\documentclass[10pt, aps, pra, twocolumn, unsortedaddress, superscriptaddress, nofootinbib]{revtex4-2}
\usepackage{graphicx} 
\usepackage{amsmath}
\usepackage{amsfonts}
\usepackage{tikz}
\usepackage{physics}
\usepackage{xcolor}

\newcommand{\rev}[1]{#1}

\renewcommand{\Re}{\text{Re}}
\renewcommand{\Im}{\text{Im}}

\begin{document}
\author{Kuntal Samanta}
\author{Sphinx J. Svensson} 
\author{Sonja Franke-Arnold} 
\email{Sonja.Franke-Arnold@glasgow.ac.uk} 
\author{Niclas Westerberg}
\affiliation{School of Physics and Astronomy, University of Glasgow, Kelvin Building, Glasgow G12 8QQ, United Kingdom}

\title{Atomic state interferometry for complex vector light}
\begin{abstract}
Features of complex vector light become important in any interference effects, including scattering, diffraction, and non-linear processes. 
Here we are investigating the role of polarization-structured light in atomic state interferometers. Unlike optical or atomic path interferometers, these facilitate local interference between atomic transition amplitudes and hence the orthogonal optical polarization components driving these transitions. We develop a fully analytical description for the interaction of generalized structured light with an atomic four state system, that is multiply connected via optical as well as magnetic transitions.  Our model allows us to identify spatially dependent dark states, associated with spatially structured absorption coefficients, which are defined by the geometry of the polarization state and the magnetic field direction.
We illustrate this for a range of optical beams including polarization vortices, optical skyrmions and polarization lattices. 
This results in a new interpretation and an enhanced understanding of atomic state interferometry, and a versatile mechanism to modify and control optical absorption as a function of polarization and magnetic field alignment. 
\end{abstract}

\maketitle


\section{Introduction}

Atomic coherence and quantum interference lie at the heart of many surprising and technically valuable effects arising from light-matter interactions. This is maybe most clearly evidenced in atomic state interferometers, where the coherent excitation of alternative transition amplitudes allows us to design and control the dielectric response. The idea of atomic state interferometers, or phaseonium, has been recognized as a resource early on \cite{Buckle1986AtomicIP,Scully1992,Kosachiov1992}, and led to effects including Electromagnetically Induced Transparency (EIT) \cite{Boller1991}, Coherent Population Trapping (CPT) \cite{Maichen1996}, lasing without inversion \cite{Harris1989,Kocharovskaya1990}, enhanced and suppressed spontaneous emission \cite{Zhu1996}, reduction and cancellation of absorption \cite{PhysRevLett.62.2813}, phase sensitive atom localization \cite{Kosachiov1993,Liu2006}. Various works have investigated phase-sensitive population dynamics theoretically \cite{Shahriar1990,Georgiades1996,Korsunsky1999,Lukin2000,Morigi2002} and through experimental observations \cite{Merriam2000,HUSS2002}.

In atomic \textit{path} interferometers (often referred to simply as atom interferometers), it is the recombination of coherent matter waves that have traveled along different paths that results in fringes of the atomic densities and coherences \cite{Keith1988,Kasevich1991,Adam1994}. Phase shifts induced by inertial forces, or through the interaction of gravitational fields, thus become measurable, making atom interferometers an ultra-precise tool for modern quantum metrology \cite{Bongs2019,HosseiniArani2024}. 

In contrast, atomic \textit{state} interferometers rely on the interference of transition amplitudes in the atomic state space within an individual atom, which is sensitive to the relative laser phase in multi-photon excitations. Atomic state interferometers can be realized, for example, via multiply connected optical transitions, such as double $\Lambda$ or diamond systems, but alternatively, states may also be coupled via microwave transitions or magnetically. In recent years atomic state interferometers have received considerable interest in order to manipulate and control light in atomic media and making it applicable to the realization of electromagnetically induced gratings \cite{Badshah2023}, atomic based microwave interferometry \cite{Shylla2018}. 

Almost all investigations of atomic state interferometers involve homogeneously polarized laser light. The incorporation of vector light beams with spatially varying polarization structures \cite{Zhan2009,Galvez2012,Rosales-Guzmán_2018,Rubinsztein-Dunlop_2017} naturally open up new avenues for atomic state interferometers. The atomic transition amplitudes are typically realized via atomic dipole transitions, which are sensitive to the alignment between the optical polarization and the electric dipole moment \cite{Wang2020,Fatemi2011,Kumar2023,Bougouffa2025}. Recent experiments based on atomic state interferometers driven by complex vector light have demonstrated their potential application for detecting the alignment of 3D magnetic fields \cite{wang2025pulsed,Castellucci2021,Qiu2021}, and theoretical work has suggested sensitivity to AC magnetic fields \cite{Ramakrishna2024}. 

Previous theoretical descriptions have been based either on solving Liouville/Bloch equations numerically \cite{Sharma2017,Ramakrishna2024,Wang2024,5r63-5hl1} or on analyzing the interactions for very specific configurations \cite{Castellucci2021,Radwell2015,PhysRevA.110.063720,Kudriasov2025}. 
Here, our goal is to provide a general analytical framework for atomic state interferometers driven by complex vector light, and to analyze the interplay between the external magnetic field, the optical polarization and the atomic spin alignment. 

Specifically, we will consider an atomic state interferometer consisting of four atomic states, with two nearly degenerate ground states coupled optically via an excited state, as well as magnetically via an intermediate ground state. We will derive an analytical model for the interaction of such phaseonium with vector light to describe and analyze its dielectric properties. We achieve this by converting the closed-loop transition into a ladder system operating on partially dressed states (which is discussed in the context of Fig. \ref{fig:atomic level scheme}). For the former, the dynamics is contained in the interplay between alternative transition amplitudes, whereas for the latter, it is reduced to a product of transition rates which can be easily evaluated in perturbation theory. 

With this method, we obtain spatially dependent dark states that no longer interact with the light and hence render the phaseonium transparent. The shapes of these dark states depend on the local polarization state of the light and its orientation with respect to the external magnetic field direction. This provides a direct link between the polarization states (as specified by its coordinates on a Poincar{\'e} sphere) and atomic transition rates. We illustrate our ideas for a range of complex vector light, ranging from vector vortex beams, to optical skyrmions \cite{Tsesses2018,Gao2020,Ye2024}, to polarization lattices.

Our study enhances the understanding of vectorial light-atom interaction, and may pave the way for encoding polarization profiles into atomic dark states, which offer protection from decoherence and noise, and for designing devices for spatially enhanced quantum magnetometry and metrology. 

The paper is organized as follows: 
In Sec. II we introduce the Hamiltonian describing the interaction of an optical field with arbitrary polarization with our phaseonium in the presence of a uniform magnetic field. We rewrite the atomic dynamics in terms of partially dressed states, that allow us to unwrap the atomic state interferometer into a ladder system. The various atomic transition rates then become functions of the polarization state and the magnetic field direction, from which we can identify the overall absorption based on perturbation theory. 
\\
Sec. III illustrates our theoretical description for various kinds of vector light beams composed of orthogonally polarized Laguerre-Gaussian modes or Hermite-Gaussian modes, including optical skyrmions. 
We finally offer our conclusions in Sec. IV.

\section{An atomic state interferometer for complex vector light}

\begin{figure}
    \centering
\includegraphics[width=0.99\linewidth]{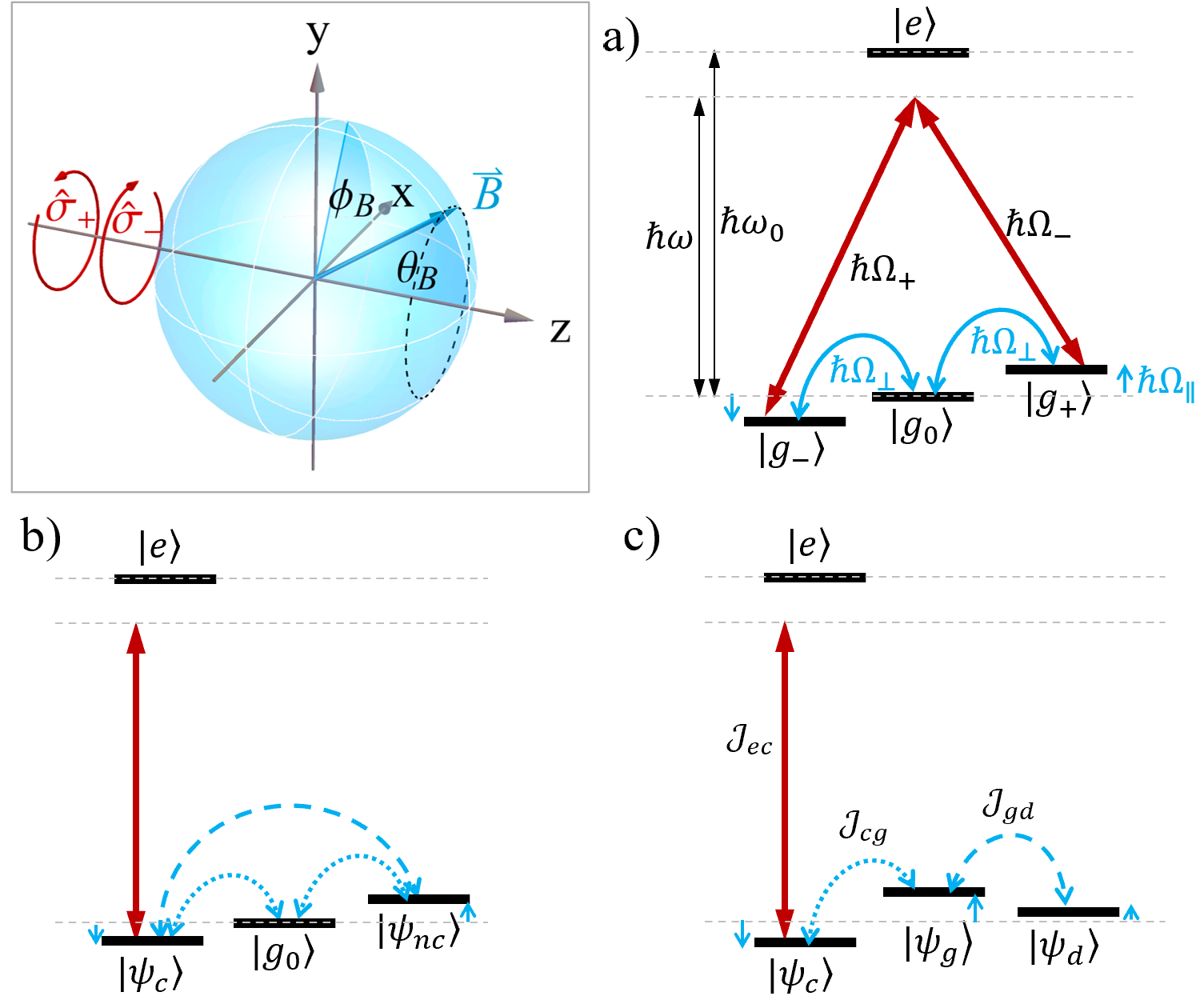}
    \caption{Schematic level scheme of the atomic state interferometer. Inset: Geometry of light propagation (along $\hat{z}$) and the magnetic field direction, defined by the azimuthal angle $\phi_B$ and the inclination angle $\theta_B$. a) Atomic state interferometer with optical and magnetic transitions. b) Intermediate partially dressed states. c) Partially dressed state systems, with optical coupling between the excited state and $\ket{\psi_c}$, with magnetic coupling driving transitions to the gray state $\ket{\psi_g}$ and from there to the dark state $\ket{\psi_d}$.}
    \label{fig:atomic level scheme}
\end{figure}

In this paper, we investigate the interaction of vector light with an atomic state interferometer, with the aim to relate the emerging internal atomic dynamics and the associated absorption and dispersion features to the properties of the vector light. As an example atomic state interferometer, we choose the optical dipole transition $F=1 \to F'=0$, driven by a quasi-resonant vectorial light-field $\vec{E}$ in the presence of a static uniform magnetic field $\vec{B}$. Specifically, we consider the electronic ground states $F=1$ with $m_F\in \{0,\pm 1\}$, and the excited state with $F=0$ and $m_F=0$, which we denote as $\ket{g_0}$ and $\ket{g_{\pm}}$ and $\ket{e}$ respectively, as indicated in Fig.~\ref{fig:atomic level scheme}a). 
Such dynamics may be realized, e.g.~by driving the \rev{$|5S_{1/2},F=1\rangle \to |5P_{3/2},F'=0\rangle$ transition in} Rb$^{87}$  F = 1 to $F^{'} = 0$ hyperfine state transition, \rev{so that interactions are restricted between the three ground states $|g_0\rangle$ ($m_F=0$) and $|g_\pm\rangle$ ($m_F=\pm 1$) and the excited state $|e\rangle$ ( $m_{F'}=0$),} but our model is applicable to any similar atomic system.

Our starting point will be the lab-frame Hamiltonian describing the atom-light interactions. Optical dipole interaction in the presence of a homogeneous external magnetic field $\vec{B}$ is given by 
\begin{align}
\hat{H}_\text{lab} & = \hbar\omega_0 \ket {e} \bra {e}+\hat{H}_\text{D} +\hat{H}_\text{Z} \nonumber\\
& =\hbar\omega_0 \ket {e} \bra {e}-\vec{d} \cdot \vec{E}(\vec{r}_A)-\vec{\mu} \cdot \vec{B}(\vec{r}_A),
\end{align}
where $\omega_0$ is the atomic resonance frequency, and $\vec{d}$ and $\vec{\mu}$ are the induced atomic electric dipole moment and atomic magnetic dipole moment, respectively. The electromagnetic fields are evaluated at the position of the atom $\vec{r}_A$, which in the following we will omit for notational simplicity. We assume that the light is propagating along the $z$-axis, which we also choose as quantization axis. We restrict ourselves to a uniform static magnetic field $\vec{B}$, but consider the interaction with a paraxial light beam whose phase, intensity and polarization may vary across its beam profile. 

\begin{figure}
    \centering
    \includegraphics[width=0.9
    \linewidth]{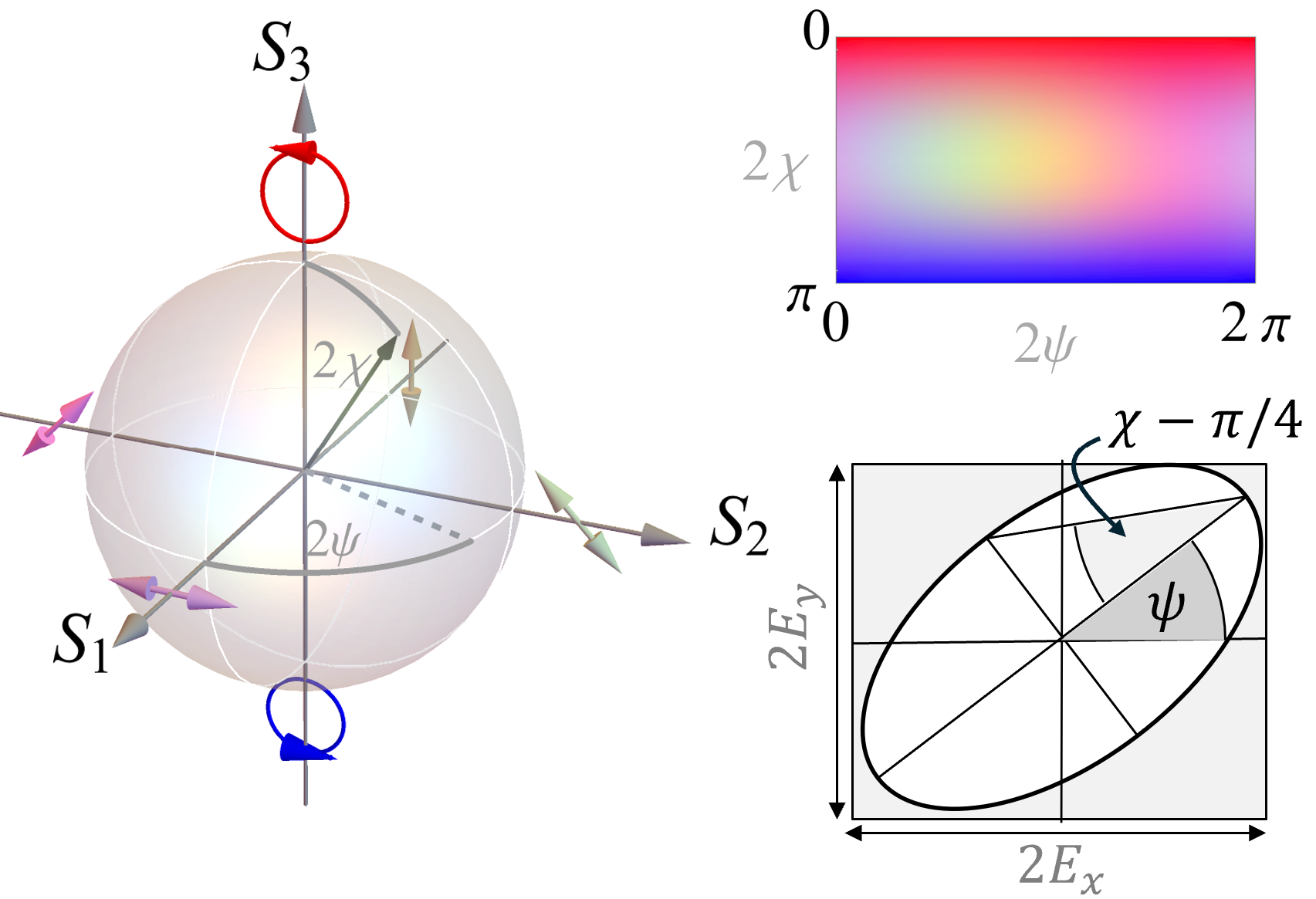}
    \caption{Definition of the optical polarization in terms of the Poincar\'e sphere and the associated polarization colour map. The spherical coordinates $\chi$ and $\psi$ uniquely define the polarization ellipse. 
    }
    \label{fig:poincare sphere}
\end{figure}

Any paraxial light beam $\vec{E}$ may be described by a two-dimensional vector
\begin{align} 
\vec{E}(\vec{r}_{\bot})&= E_0(\vec{r}_{\bot}) e^{i\omega t} \left[ u_+ (\vec{r}_{\bot}) \hat{\sigma}  _+ +u_-(\vec{r}_{\bot}) \hat{\sigma}  _-\right] + \text{c.c}. 
\end{align}
which varies as a function of the spatial coordinate $\vec{r}_\bot = (x,y)$ within the beam profile. As atom transitions are expressed in terms of $\sigma_\pm$ transitions, we have however decomposed the light field into its left and right polarization components $\hat{\sigma}_{\pm}=\left( \hat{x}\pm i\hat{y} \right)/\sqrt{2}$, driving these transitions. The circular polarization components are associated with spatially varying complex amplitudes $u_\pm$, with $|u_+ (\vec{r}_{\bot})|^2+ |u_- (\vec{r}_{\bot})|^2=1$. 
Without restriction of generality, we may decompose the electric field into a common total complex amplitude, $E_0(\vec{r}_{\bot})$, and a polarization state, by expanding the complex amplitudes as
\begin{equation}
    u_- = { \cos{\chi} } \; e^{-i \psi},\quad
    u_+ = { \sin{\chi} } \; e^{i \psi},
\end{equation}
so that the electric field becomes
\begin{align} \label{eq:E-field}
\vec{E}(\vec{r}_{\bot})&=  E_0(\vec{r}_{\bot}) e^{i\omega t} \left[ { \sin \chi(\vec{r}_{\bot}) }\,e^{i\psi(\vec{r}_\perp)} \hat{\sigma}  _+\right. \\
&\left.\hspace{2.2cm}+  \cos \chi(\vec{r}_{\bot}) \,e^{-i\psi(\vec{r}_{\bot})} \hat{\sigma}  _-\right] + \text{c.c.}  \nonumber
\end{align}
In this notation, the polarization state is parametrized by the parameters $2\chi$ and $2\psi$ describing the degree of ellipticity and orientation of the polarization ellipse respectively. 
They can be understood as the spherical coordinates of a unique point on the Poincar{\'e} sphere, as illustrated in Fig.~\ref{fig:poincare sphere}, where $2\chi$ denotes the polar angle as measured from the North pole, while $2\psi$ represents the azimuthal angle. 
This definition relates the polarization state to the (local) reduced Stokes parameters
\begin{align} 
\vec{S} = \begin{pmatrix}
S_1 \\
S_2 \\ 
S_3
\end{pmatrix} =  \begin{pmatrix}
\sin{2\chi} \,\cos2\psi\\
\sin{2\chi}\,\sin2\psi \\ 
\cos{2\chi}
\end{pmatrix} =   \begin{pmatrix}
2 \Re(u_+^* u_-)\\
-2 \Im(u_+^* u_-) \\ 
| u_- |^2-| u_+ |^2
\end{pmatrix}.
  \label{eq:stokes} 
\end{align}
We note that our definition is equivalent to the more common definition in terms of an imbalance of horizontal to vertical, diagonal to anti-diagonal and right to left circular polarization components. 

We then express the homogeneous magnetic field \rev{in polar coordinates} as
\begin{align}
\vec{B} &= B_0 \left(\cos\theta_B \hat{z} + \sin\theta_B \cos{\phi_B} \hat{x}-\sin\theta_B \sin\phi_B \hat{y} \right) \nonumber\\
& = B_0 \left( \cos\theta_B \hat{z} + \sin\theta_B\frac{\left[ e^{-i\phi_B} \hat{\sigma}_- + e^{i\phi_B} \hat{\sigma}_+ \right]}{\sqrt{2}}  \right),
\end{align}
where $\theta_B$ is the tilt angle between the magnetic field and the propagation axis $\hat{z}$, and $\phi_B$ is the azimuthal angle, measured in clockwise direction from the vertical, as indicated in the inset of Fig.~\ref{fig:atomic level scheme}. The second equality is in terms of cylindrical coordinates. The magnetic dipole moment then becomes
\begin{align*}
\vec{\mu}&=-g_F \mu_B \left[ \hat{z} \left( \ket{g_+} \bra{g_+} -\ket{g_-}\bra{g_-} \right) \nonumber\right.\\
&\left. \hspace{1.5cm}-\left(\hat{\sigma}_- \ket{g_+}\bra{g_0} +\hat{\sigma}_+ \ket{g_-}\bra{g_0} + \text{H.c.}\right) \right],
\end{align*}
where the terms $g_F$ and $\mu_B$ are the Land{\'e} g-factor and Bohr magneton, respectively. Introducing the Larmor frequency as $\Omega_L = g_F \mu_B B_0 / \hbar$, we write the magnetic dipole interaction as
\begin{align}
\hat{H}_Z&=-\vec{\mu}\cdot \vec{B}\\
&=\hbar \Omega_L \left[ \cos\theta_B \left( \ket{g_+} \bra{g_+}-\ket{g_-} \bra{g_-} \right)\right.\nonumber \\
&\left.\hspace{.8cm}-\sin\theta_B \left( e^{i\phi_B} \ket{g_+}\bra{g_0} +e^{-i\phi_B} \ket{g_-} \bra{g_0}\right)/\sqrt{2}  \right],\nonumber
\end{align}
which includes the Zeeman-shift due to the magnetic field component along the quantization axis as well as magnetic coupling due to its transverse components.
 
We assume that the light is paraxial, so that any component of the optical field along the propagation direction is negligible, and we can ignore any excitation of the $\pi$-transition from $\ket{g_0}$ to $\ket{e}$. The only relevant parts of the electric dipole moment are therefore 
\begin{equation}
 \hat{\vec{d}}=d\left[ \hat{\sigma}_+ \ket{e}\bra{g_+}+\hat{\sigma}_- \ket{e} \bra{g_-}\right]/2 \sqrt{3} +\text{H.c.},
\end{equation}
yielding the optical dipole Hamiltonian
\begin{align}
\hat{H}_D &= -\hat{\vec{d}} \cdot \vec{E} \\
&=-\frac{\hbar\Omega_R}{2 \sqrt 3} \left[ e^{i \psi}\left({ \cos{\chi} }\,  e^{-i\omega t} +{ \sin{\chi} }\, e^{i\omega t} \right) \ket{g_+} \bra{e} \right. \nonumber\\ 
&\hspace{0.5cm} + \left. e^{-i \psi} \left({\sin{\chi} }\,  e^{-i\omega t} + { \cos{\chi} }\,  e^{i\omega t} \right)\ket{g_-}\bra{e} \right] + \text{H.c.},\nonumber
\end{align}
where $\Omega_R = d E_0$ denotes the Rabi frequency and the factor $1/\sqrt{3}$ originates from the Wigner-Eckart coefficients. In the frame co-rotating\footnote{This is done by the unitary transform,
$\hat{H_I}=\hat{U}\hat{H}\hat{U}^{\dagger}+i\frac{\partial U}{\partial t} \hat{U}^{\dagger}$, where $\hat{U}=\exp\left[i\omega t \ket{e} \bra{e}\right]=\ket{g_+} \bra{g_+}+\ket{g_0} \bra{g_0}+\ket{g_-}\bra{g_-}+\ket{e} \bra{e} e^{i\omega t}$} with the electric field, the Hamiltonian becomes
\begin{align}
\hat{H}/\hbar
&=-\delta\ket{e}\bra{e}+\Omega_L \cos\theta_B\left(\ket{g_+} \bra{g_+}-\ket{g_-}\bra{g_-}\right) \nonumber\\
&\hspace{0cm} -\frac{\Omega_R}{2\sqrt{3}}\left[{ \sin{\chi} }\, e^{i \psi}\ket{g_+}\bra{e}+{ \cos{\chi} }\, e^{-i \psi} \ket{g_-}\bra{e} + \text{H.c.} \right]  \nonumber\\
&\hspace{0cm}-\frac{\Omega_L}{\sqrt{2}}\sin\theta_B\left[e^{i\phi_B}\ket{g_+}\bra{g_0}+e^{-i\phi_B}\ket{g_-}\bra{g_0}+\text{H.c.} \right], 
\end{align}
for the detuning $\delta=\omega-\omega_0.$
This is the complete Hamiltonian in the rotating wave approximation, expressed in terms of the atomic states $\ket{g_-},\ket{g_0},\ket{g_+}$ and $\ket{e}$. Its first line denotes the energies, including the Zeeman shift,  the second line describes Rabi oscillations due to optical coupling, and the third Larmor precession due to magnetic coupling. The action of the Hamiltonian is schematically indicated in Fig.~\ref{fig:atomic level scheme}(a). Any two states within this atomic state interferometer are coupled via two alternative electric and/or magnetic transition amplitudes, which can interfere. The resulting dynamics, and specifically the absorption and dispersion, therefore should depend on the differential phase between the optical transitions and on the alignment of the magnetic field.
The dynamics can be determined numerically by solving the corresponding Liouville or Bloch equation, including decay and relaxation rates.  
The evaluation of the Liouville or Bloch equation is, however, computationally intensive, prohibiting a comprehensive investigation of the parameter space of arbitrary magnetic fields and polarizations, and its evaluation does not lead to an intuitive understanding of the system geometries.

\subsection{Hamiltonian in terms of partially dressed states}
In this work, we pursue a different approach, leading to a fully analytical solution of populations and transition rates. The aim is to transform the interfering transition amplitudes, i.e.~coherences, of the $\hat{\sigma}_\pm$-transitions into population dynamics of a ladder-like system. The interference is then captured in the form of partially dressed states and the emergent hopping rates.
To achieve this, we will apply unitary transforms in such a way that the atomic state interferometer is unwrapped into a ladder system, as indicated in Fig.~\ref{fig:atomic level scheme}c). The partially dressed states will comprise a single coupling state $\ket{\psi_c}$, connected by optical transitions to the excited state, as well as two other partially dressed states which we will term gray state $\ket{\psi_g}$ and dark state $\ket{\psi_d}$ (note that the latter is not a true dark state, but for certain geometries can become one). 
The form of the optical coupling suggests the introduction of a coupling state $\ket{\psi_c}$ and its orthogonal non-coupling state $\ket{\psi_{nc}}$:
\begin{align}
 \label{eq:Coupling state} 
\ket{\psi_c}={ \sin{\chi} }\, e^{i \psi} \ket{g_+}+ { \cos{\chi} }\, e^{-i \psi}\ket{g_-}, \\
\label{eq:Non-coupling state} 
\ket{\psi_{nc}}= { \cos{\chi} }\, e^{i \psi}\ket{g_+}-{ \sin{\chi} }\, e^{-i \psi}\ket{g_-}.
\end{align}

The Hamiltonian can then be rewritten as
\begin{widetext}
\begin{align}
\label{eq:H_C_NC}
\frac{\hat{H}}{\hbar}~&=~-\delta\ket{e}\bra{e}+\Omega_L \cos\theta_B\left[ -\cos{2\chi}\left( \ket{\psi_c}\bra{\psi_c}-\ket{\psi_{nc}} \bra{\psi_{nc}}\right)+\left(\sin{2\chi}\ket{\psi_c}\bra{\psi_{nc}}+\text{H.c.}\right)\right]\nonumber\\
  &\hspace{1cm} - \frac{\Omega_L}{\sqrt{2}} \sin\theta_B\bigg[e^{-i(\psi-\phi_B)\,}{ \sin{\chi} }\, \ket{\psi_c}\bra{g_0}+ { \cos{\chi} }\, e^{-i(\psi-\phi_B)}\ket{\psi_{nc}}\bra{g_0}\nonumber\\
  &\hspace{1cm} + e^{i(\psi-\phi_B)} { \cos{\chi} }\, \ket{\psi_c}\bra{g_0}-{ \sin{\chi} }\, e^{i (\psi-\phi_B)}\ket{\psi_{nc}}\bra{g_0} +\text{H.c}\bigg]
  -\frac{\Omega_R}{2\sqrt{3}}\left(\ket{\psi_c}\bra{e}+\text{H.c}\right).
\end{align}
\end{widetext}

In this expression we recognize the factors $S_3=\cos{2\chi}$ and $\sqrt{S_1^2+S_2^2}=\sin{2\chi}$ from the definition of the Stokes vectors Eq.~(\ref{eq:stokes}). 
We note that in Eq.~\eqref{eq:H_C_NC} the azimuthal angle $\psi$ and the azimuthal angle of the magnetic field $\phi_B$ always appear in combination with each other. This is not surprising, as the interaction is set by the geometry of the system, {\it i.e.}~the angle of the local polarization direction against the magnetic field direction. 
In the following we will denote their difference as 
\begin{equation} \label{eq_phiB} \psi'=\psi-\phi_B.
\end{equation}
For notational simplicity, let us introduce two complex 
parameters, $J$ and $\bar{J}$, as

\begin{align}
\begin{split}
  J &= \frac{1}{\sqrt{2}}\left[ e^{i \psi'} { \cos{\chi} }\, +e^{-i\psi'}{ \sin{\chi} }\,\right],\\ 
  \bar{J} &= \frac{1}{\sqrt{2}}\left[- e^{i \psi'} { \cos{\chi} }\, +e^{-i \psi'}{ \sin{\chi} }\,\right].
\end{split}
\end{align}

The Hamiltonian now takes the form
\begin{align}
\frac{\hat{H}}{\hbar}=&-\delta \ket{e}\bra{e}-\Omega_L\cos\theta_B \cos{2\chi} \left(\ket{\psi_c} \bra{\psi_c}-\ket{\psi_{nc}}\bra{\psi_{nc}} \right)\nonumber\\
&+\Omega_L\left[\cos\theta_B \sin{2\chi}\ket{\psi_c}\bra{\psi_{nc}} -\sin\theta_B J \ket{\psi_c}\bra{g_0} \right.\nonumber\\
&\hspace{2.5cm} \left. +\sin\theta_B \bar{J}\ket{g_0}\bra{\psi_{nc}} + \text{H.c.}\right] \nonumber \\
& -\Omega_R \big(\ket{\psi_c}\bra{e}+\text{H.c.}\big)/2\sqrt{3}.\label{eq:part state H 1}
\end{align}
Whilst this form allows only one state ($\ket{\psi_c}$) to couple to the light, it nonetheless contains also two magnetically driven transitions from the coupling state, as depicted in Fig.~\ref{fig:atomic level scheme}.b), effectively forming an atomic state interferometer within the ground states. Moreover, the hopping rates between $\ket{g_0}$ and $\ket{\psi_c}$, and between $\ket{g_0}$ and $\ket{\psi_{nc}}$ are generally complex (unless the polarization is exactly aligned or perpendicular with the transverse magnetic field when $\psi'=\psi-\phi_B=0$), indicating a directional transition direction between the ground states. 

We can remove this feature by, once more, rewriting the Hamiltonian in terms of new system states generated from superpositions of the non-coupling state ($\ket{\psi_{nc}}$) and $\ket{g_0}$. As we shall see, these states have physical significance when considering the atomic dynamics, in other words: we now reach the aforementioned gray state $\ket{\psi_g}$ and dark state $\ket{\psi{_d}}$.

These, together with the previously defined coupling state $\ket{\psi_{c}}$ from Eq.~(\ref{eq:Coupling state}) give our final orthonormal basis set of partially dressed ground states:
\begin{align}
\ket{\psi_c} & ={ \sin{\chi} }\, e^{i \psi} \ket{g_+}+ { \cos{\chi} }\, e^{-i \psi}\ket{g_-},\nonumber\\
\ket{\psi_g} &= \frac{1}{M}\left(\cos\theta_B \sin{2\chi} \ket{\psi_{nc}}-\sin\theta_B J^*\ket{g_0} \right), \label{eq:states}\\
\ket{\psi_d} &= \frac{1}{M}\left(\sin\theta_B J\ket{\psi_{nc}}+\cos\theta_B \sin{2\chi} \ket{g_0}\right),\nonumber
\end{align}
where $M$ is a normalization constant defined by 
\begin{align} 
\label{eq:normalization_const}
M^2=(1-\cos^2{\theta_B \cos{4\chi} +\cos{2\psi'} \sin^2{\theta_B} \sin{2\chi)/2}}.
\end{align}

\begin{figure*}[ht]
    \centering
\includegraphics[width=\linewidth]{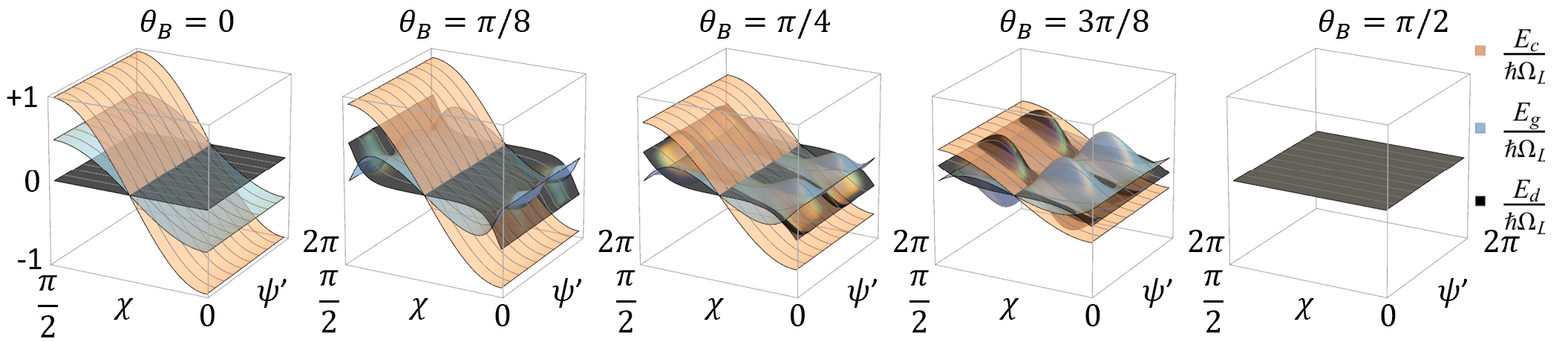}
    \caption{Energies of the partially dressed ground states $|\psi_c\rangle$, $|\psi_g\rangle$, and $|\psi_d\rangle$ as a function of the optical polarization and magnetic field orientations. The polarization states show the energies for a magnetic field that tilted with respect to the propagation direction by an angle $\theta_B$ between 0 and $\pi/2$.}
    \label{fig:Energy_combined}
\end{figure*}

Note that the value of $M$ depends on the inclination of the magnetic field with respect to the propagation direction, and on the polarization state, while the parameters $J$ and $\bar J$ depend solely on the optical polarization. Expressed in terms of the states in Eq.~(\ref{eq:states}), the desired Hamiltonian, represented in Fig.~\ref{fig:atomic level scheme}.c), becomes:
\begin{align} \label{eq:HamiltonianGeneric}
\hat{H} &= E_c\ket{\psi_c}\!\bra{\psi_c} + E_g\ket{\psi_g}\!\bra{\psi_g} + E_d\ket{\psi_d}\!\bra{\psi_d} -\hbar\delta\ket{e}\!\bra{e}\nonumber \\
& + \mathcal{J}_{ec}\ket{e}\bra{\psi_c}+ \mathcal{J}_{cg}\ket{\psi_c}\bra{\psi_g} + \mathcal{J}_{gd}\ket{\psi_g}\bra{\psi_d} + \text{H.c.}
\end{align}
The first line of this Hamiltonian contains the energies of the coupling, gray, dark and excited states, where we have defined
\begin{align}
    E_c &= -\hbar\Omega_L\cos\theta_B \cos{2\chi},  \label{eq:energies}\\
    E_g &= -\hbar\Omega_L\frac{\cos\theta_B\sin{4\chi}}{2 M^2} \Big(\cos^2\theta_B  \sin{2\chi}   \nonumber\\ 
&\hspace{4cm}+\sin^2\theta_B\cos2\psi'\Big),\nonumber\\
    E_d &= -\hbar\Omega_L\frac{\cos\theta_B\sin^2\theta_B}{2 M^2} \cos{2\chi} \left({1}-{\sin{2\chi} }\cos2\psi'\right). \nonumber
   \end{align}
The dependence of these energies on optical polarization and magnetic field orientations is depicted in Fig.~\ref{fig:Energy_combined}.
As the composition of our system states Eq.~(\ref{eq:states}) is set by the configuration of electric and magnetic field, their Zeeman shifts are no longer just defined by the longitudinal magnetic field component, but also the ellipticity of the incident light field and the alignment between the transverse magnetic field and the orientation of the polarization ellipse.

The second line of Eq.~(\ref{eq:HamiltonianGeneric}) describes transitions between the states, where the hopping rates are given by
\begin{align}
\label{eq:J}
   \mathcal{J}_{ec} &= -\hbar\Omega_R/2\sqrt{3}, \nonumber \\
    \mathcal{J}_{cg} &= \hbar M \Omega_L/2, \\
    \mathcal{J}_{gd} &= -\hbar\Omega_L\frac{\bar{J}}{M^2}\sin\theta_B\left(\sin{2\chi}  \cos^2\theta_B+\sin^2\theta_B J^2\right). \nonumber
\end{align}
Here $\mathcal{J}_{ec}$ and $\mathcal{J}_{cg}$ are real, while $\mathcal{J}_{gd}$ is generally complex but becomes real for $\psi'=0$.

We have thus reached a ladder-form, depicted in Fig.~\ref{fig:atomic level scheme}c), where each state only couples only with one other state: excited to coupling state, coupling to gray state, and finally gray to dark state. 
Both $\mathcal{J}_{cg}$ and $\mathcal{J}_{gd}$ depend on the geometry between the magnetic field direction and the local polarization states.
 
If the coupling vanishes for any configuration of electric and magnetic fields, $\ket{\psi_d}$ becomes a dark state, which will be filled rapidly by spontaneous decay from the excited state. Similarly, for any configuration with $\theta_B=n\pi$, $\mathcal{J}_{cg}$ vanishes, making both $\ket{\psi_g}$ and $\ket{\psi_d}$ into dark states. For these parameters, light will no longer be absorbed but can pass unhindered through the atomic sample.

For uniformly polarized light fields, the atomic interaction will (up to saturation effects) be uniform across the beam profile, but vary as a function of the alignment of the magnetic field and the chosen polarization state. For vector light, with spatially varying polarization profiles, instead, absorption (and also dispersion) will be modulated across the beam profile, resulting in spatially dependent EIT. Such vector light provides intriguing possibilities to explore the relationship between the various energy and hopping rates and the geometries of the optical polarization and the magnetic field direction, specified by the Stokes parameters and the magnetic field angles $\theta_B$ and $\phi_B$.

\begin{figure*}[ht]
    \centering
\includegraphics[width=\linewidth]{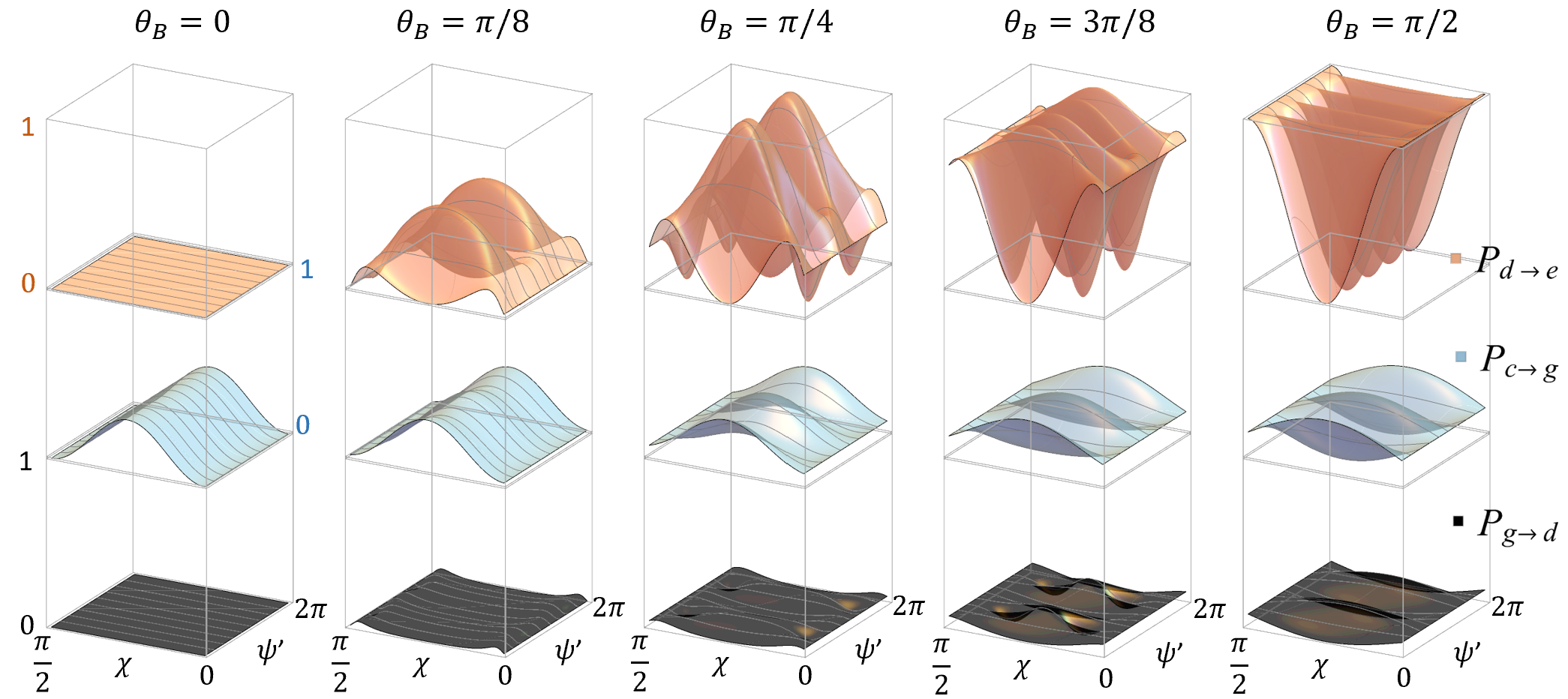}
    \caption{Absorption rate, proportional to the probability for an atom to transition between the dark and excited state as a function of the optical polarization and magnetic field inclination. The first row shows $P_{d\to e}$, which is a product of the probabilities $P_{g\to c}$ (second row), $P_{d\to g}$ (third row), and $P_{c\to e}$ (independent of polarization and magnetic field angle, not shown. All transition rates are peak normalised for $P_{d\to e}$. Here we assume $\Omega_L\ll\Gamma$.}
    \label{fig:Transitions_combined}
\end{figure*}

\subsection{The absorption rate}
We are interested in calculating the probability for an electron to move from the dark to the excited state, 
\begin{align}
    P_{d\rightarrow e}~=~\left|\bra{e}\ket{\psi_d(t)}\right|^2,
\end{align}
as this is a measure of the electric field absorption \cite{Radwell2015, Wang2024}. Following a perturbative approach outlined in Appendix~\ref{app:pertTheory}, we find the transition probability 
\begin{align} \label{Eq:P}
    P_{d \rightarrow e} &\simeq \frac{1}{36\Gamma^6}\left|\mathcal{J}_{ec} \mathcal{J}_{cg} \mathcal{J}_{gd}\right|^2 \nonumber\\
    &= \frac{\Omega_R^2}{432\hbar^4\Gamma^6}\left|\mathcal{J}_{cg} \mathcal{J}_{gd}\right|^2,\nonumber \\
     &=  \frac{\Omega_R^2 \Omega_L^2}{1728\hbar^2\Gamma^6} M^2 \left|\mathcal{J}_{gd}\right|^2, 
\end{align}
where we have used the expressions for $\mathcal{J}_{ec}$  and $\mathcal{J}_{cg}$ from Eq.~\eqref{eq:J}. This is the central result of this manuscript. We have here introduced the lifetime of the excited state $\Gamma$, which we assume to be sufficiently large as compared to $\Omega_R$ and $\Omega_L$ as to ensure that the physics is well-captured by the short time dynamics. 

It is worth noting, that $\ket{\psi_d}$ is truly dark only for specific parameters determined by the local Stokes angles of the polarization and the magnetic field alignment. 
For an atom to undergo a transition from the dark state $\ket{\psi_d}$ to the excited state $\ket{e}$, it must progress along the transition ladder via the gray state $\ket{\psi_g}$ to the coupling state $\ket{\psi_c}$ before it can be optically excited. As the optical transition rate $|\mathcal{J}_{gd}|^2$ is spatially homogeneous, the geometry of the transition rate $P_{d \rightarrow e}$ is therefore determined by the product of the transition rates $P_{d \rightarrow g}$ and $P_{g \rightarrow c}$, Alternatively, given the simple form of $\mathcal{J}_{cg}$ in Eq.~\eqref{eq:J}, the geometrical factors can be found from $ M^2 \left|\mathcal{J}_{gd}\right|^2$. 
 
The absorption rate as a function of $\chi$ and $\psi'$ is illustrated in Fig.~\ref{fig:Transitions_combined}. The transition rates depend only on the \emph{alignment} of the optical polarization with respect to the magnetic field, not on its \emph{orientation} which is defined with respect to some external reference frame, hence we show only magnetic fields for $0\leq \theta_B\leq\pi/2$. For the same reason we restrict ourselves to presenting polarizations corresponding to the `northern hemisphere` of the Poincar\'e sphere (i.e.~light with an angular momentum along the $z$ direction), which acts identical to its counterpart on the `southern hemisphere' (with an angular momentum along negative $z$). Zero absorption, i.e.~true dark states correspond to polarization states for which $P_{g\rightarrow d} \propto \mathcal{J}_{gd}$ vanishes. This happens where the major or minor axis of the polarization ellipse is aligned with the magnetic field direction, i.e. for $\psi-\phi_B= n \pi/2 $ for $n \in \mathbb{N}$. In a way, the system now acts like a higher order polariser, selecting orthogonal polarisation states but not their superpositions.
The absorption patterns can be determined by multiplying the transition amplitude with the intensity profile of the corresponding beam as illustrated with a few examples in the following section.

\section{Examples}
In this section, we will illustrate our theoretical model by evaluating the dynamics for  different configurations of external magnetic fields and vector light fields, and present the predicted absorption patterns from Eq.~\eqref{Eq:P}.  Where instructive, we also provide the equations for the energies Eq.~\eqref{eq:energies} and hopping rates
Eq.~\eqref{eq:J}. 

\subsection{Polarization vortices with varying ellipticity}  
\begin{figure*}[ht]
    \centering
\includegraphics[width=1.0\textwidth]{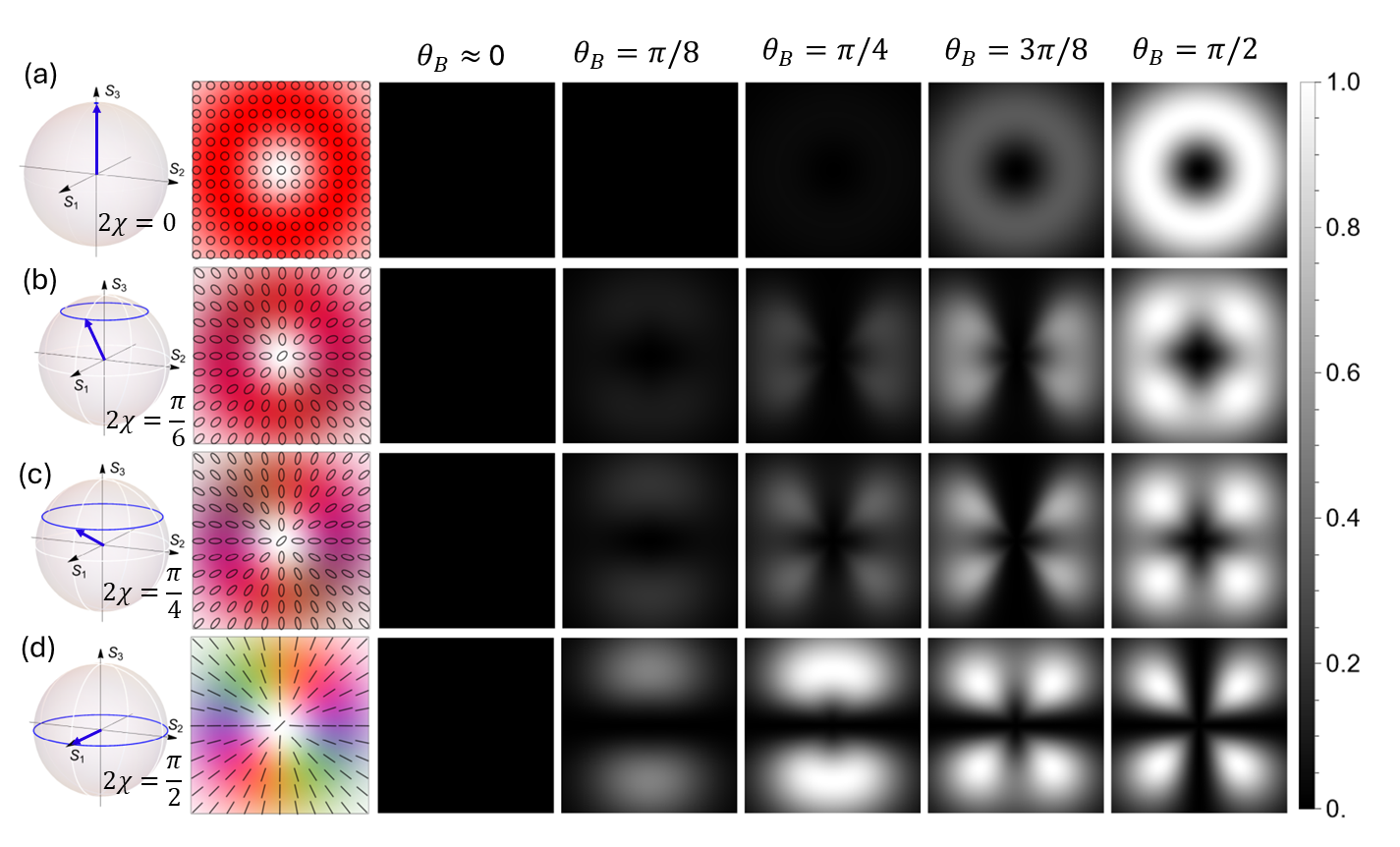}
      \caption{Absorption patterns for light with a variety of polarization profiles with varying ellipticity ranging from homogeneous right polarization in row a) to radial polarization in row d) 
      for various magnetic field inclinations $\theta_B$ (for $\phi_B =0$ so that the magnetic field rotates in the $x$-$z$ plane). Each row shows from left to right the polarizations on the Poincar\'e sphere with the arrow indicating $\varphi=0$, the corresponding beam profile, followed by the absorption patterns for $\theta_B\approx 0 $,  $\theta_B=\pi/8$, $\theta_B=\pi/4$, $\theta_B=3\pi/8$ and  $\theta_B=\pi/2$ respectively. White indicates maximum absorption.
     }
    \label{fig:concurrence}
\end{figure*}

Let us start by exploring the interaction of polarization vortices as a function of their degree of ellipticity and the inclination of the magnetic field $\theta_B$. Here we consider a family of beams with azimuthally varying orientation (i.e.~azimuthally varying $\psi$ values), parametrized by their degree of ellipticity (i.e.~the $\chi$ value). The electric field is given by
\begin{align} \label{Eq_chi}
    \vec{E}_\chi&=  E_0(r) \left[ { \sin \chi }\,e^{-i\varphi} \hat{\sigma}_+ + { \cos \chi }\,e^{i\varphi} \hat{\sigma}_-\right] + \text{c.c.}\\
    &\propto  \sin \chi \, {\rm LG}_0^{-1}  \hat{\sigma}_+ + \cos \chi \, {\rm LG}_0^{+1}  \hat{\sigma}_- +\text{c.c.,} \nonumber
\end{align}
where $\varphi$ denotes the azimuthal phase. The second line specifies a potential realization of such light in terms of different Laguerre-Gauss modes imprinted onto the opposite circular polarization components. The $\hat{\sigma}_{\pm}$ component of the light is associated with a helical phase structure $e^{\mp i \varphi}$ which denotes an orbital angular momentum (OAM) of $\mp \hbar$ per photon, and the amplitude between the opposite circular light components is controlled by the parameter $\chi$. Modifying $\chi$ from 0 to $\pi/4$ changes the beam profile from homogeneous right hand circularly polarized light to radial, with the Poynting vector rotating twice around a circle of equal latitude as a function of $\varphi$ within the beam profile, as shown in the first two columns of  
Figure \ref{fig:concurrence} for $2\chi=0,\,\pi/6, \,\pi/4$, and $\pi/2$. The associated absorption patterns can be calculated from Eq.~\eqref{Eq:P} and are shown for varying values of $\theta_B$ in the following 5 columns of this figure. 
The energies of the partial dressed states simplify to
\begin{align}
\label{energy_concurrence}
  E_c &= -\hbar\Omega_L\cos\theta_B \cos{2\chi},\nonumber \\
    E_g &= -\hbar\Omega_L\frac{\cos\theta_B\sin{4\chi}}{2 M^2} \Big(\cos^2\theta_B  \sin{2\chi}   \\ 
&\hspace{4cm}+\sin^2\theta_B\cos2\varphi'\Big),\nonumber\\
    E_d &= -\hbar\Omega_L\frac{\cos\theta_B\sin^2\theta_B}{2 M^2} \cos{2\chi} \left({1}-{\sin{2\chi} }\cos2\varphi'\right). \nonumber
\end{align}

The transition rate between dark and gray state becomes
\begin{align}
\label{hoppingrates_atomic_compass}
    \mathcal{J}_{gd} 
    &=-\frac{\hbar \Omega_L } {M^2} \sin \theta_B \big[\bar{J} \sin 2\chi \cos^2\theta_B \\
    &
   \hspace{1.0cm} -J \sin^2\theta_B \big(\cos 2\varphi' \cos2\chi
   +i \sin 2\varphi' \big)/2\big] \nonumber,
\end{align}
where 
$
M^2=(1-\cos^2\theta_B \cos4\chi+\cos 2\varphi' \sin^2\theta_B \sin2\chi)/2$, 
and we have once again introduced $\varphi'=(\varphi-\phi_B)$, as the difference between the azimuthal angle of the beam profile $\varphi$ and the azimuthal angle of the magnetic field $\phi_B$. And, as mentioned earlier, the total transition rates between the dark and excited state can be calculated from Eq.~\ref{Eq:P}, which becomes a function of $|\mathcal{J}_{gd}|^2M^2$.

We note that, generally, the visibility of the absorption pattern increases with the ellipticity angle $\chi$. 

In the following we will analyze two specific cases of the beams described by Eq.~\eqref{Eq_chi}, which have been investigated experimentally in \cite{Castellucci2021} and \cite{Wang2024}.

By setting $\chi= \pi/4$, corresponding to radially polarized light, we recover the results of Ref.~\cite{Castellucci2021}.  
In this case, all energies vanish ($E_c  = E_g = E_d  = 0$), and the hopping rate between gray and dark state is
\begin{align}
    \mathcal{J}_{gd} &=  \frac{ i\hbar \Omega_L}{4 M^2} \sin\theta_B\sin\varphi^{\prime} \big[\cos^2\theta_B+\sin^2\theta_B \cos^2\varphi^{\prime}\big],   
\end{align}
where $
M^2=(1+\cos^2\theta_B +\cos 2\varphi' \sin^2\theta_B )/2$.
 
While here we use slightly different definitions for the coupling, gray and dark states than in \cite{Castellucci2021}, the physical predictions are identical.
Our analysis here confirms the previous experimental demonstration that the magnetic field direction can be inferred from the absorption pattern: a tilt $\theta_B$ of the magnetic field changes the petal structure of the absorption pattern, and rotating the magnetic field by $\phi_B$ results in a rotation of the petal pattern as $\phi_B = \psi-\psi'$. 

On the other hand, if we let $\chi$ vary and instead assume that the magnetic field is uniform and orthogonal to the propagation axis (e.g. by setting  $\theta_B=\pi/2$ and $\phi_B=0$ which correspond to a magnetic field $\vec{B} = B_0 \hat{x}$), then we recover the results of Ref.~\cite{Wang2024}, corresponding to the final column of Fig.~\ref{fig:concurrence}. In this geometry, all Zeeman splitting disappears so that $E_c = E_g = E_d = 0$, and furthermore by definition $\psi'=\psi$. 
The hopping rate between gray and dark state is then
\begin{align}
\mathcal{J}_{gd} &= \frac{\hbar\Omega_L}{2 M^2} J({ \cos{2\chi} }\,\cos 2\psi+i\sin2\psi)
\end{align}
where $M^2=(1+\cos2\psi\, \sin 2\chi)/2$.
The degree of ellipticity $\cos2\chi$ determines whether $\mathcal{J}_{gd}$ has a real component or not. The fringe visibility of the absorption pattern allows us to determine the correlations in the polarization structure, i.e. the concurrence, as was experimentally confirmed in Ref.~\cite{Wang2024}. We note, however, that this simple correspondence breaks down for magnetic fields with $\theta_B\neq \pi/2$, showing that the relation between absorption patterns and optical concurrence is affected by the magnetic field direction - maybe not surprising given that the latter determines magnetic couplings and energy shifts within the atomic state interferometer.

\subsection{Hybrid vector beams along different grand circles} 

\begin{figure*}[ht]
   \centering
\includegraphics[width=\textwidth]{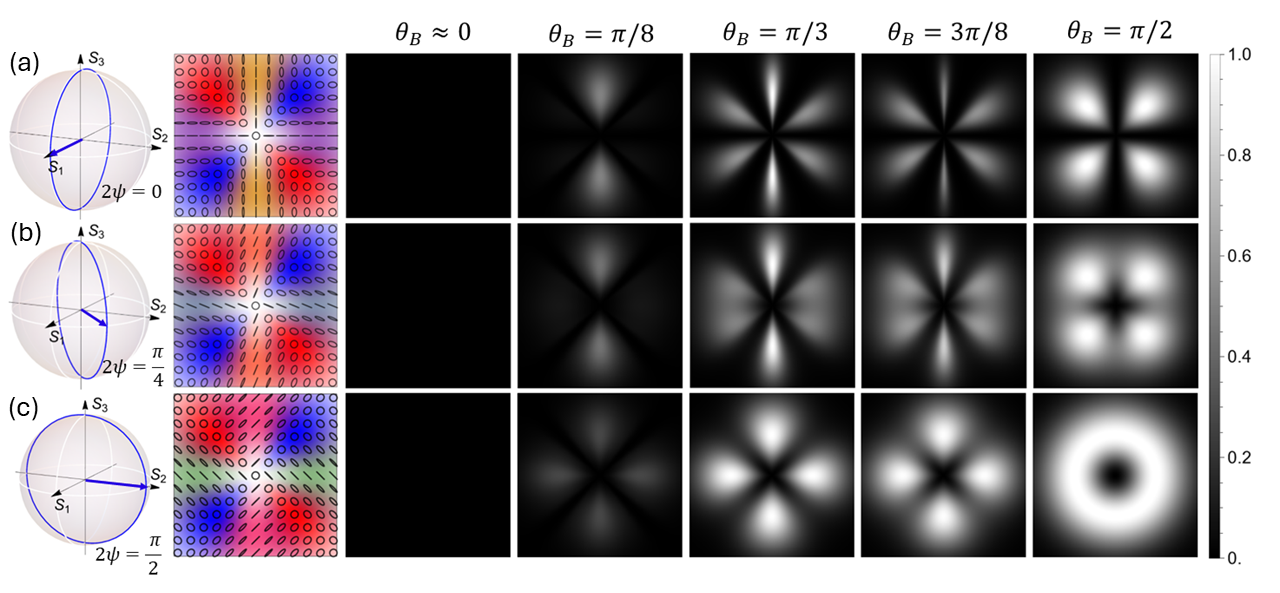}
     \caption{
     Absorption patterns for hybrid polarized light beams along differently oriented grand circles on the Poincar\'e sphere, for the same magnetic field parameters as in Fig.~\ref{fig:concurrence}, i.e. $\vec{B}=B_0 \hat{x}$. The absorption patterns display a variety of symmetries, including two-fold, four-fold, six-fold and cylindrical. (Note that the angles of $\theta_B$ do not progress linearly in order to incorporate the value of $\theta_B=\pi/3$) which shows the clearest 6-fold symmetry.}
   \label{fig:hyb concurrence}
\end{figure*}
In this section we investigate the influence of polarization alignment, determined by $\chi$, on the absorption behavior of our atomic state interferometer. We can obtain differently structured hybrid-polarized beams from Eq.~\eqref{eq:E-field} by mapping $\chi$ to the azimuthal angle $\varphi$ and varying $2\psi$ from $0$ to $\pi/2$ on the Poincar\'e sphere in Fig.~\ref{fig:poincare sphere}.  
We choose an electric field of the form
\begin{align}
\label{eq:hybrid polarization}
\vec{E}_\text{hyb}(\vec{r}_{\bot})
&= E_0(\vec{r}_{\bot}) e^{i\omega t} \left[ { \sin\, (\ell \varphi) }\,e^{i\psi} \hat{\sigma}  _+\right. \\
&\left.\hspace{2.5cm}+{ \cos\, (\ell \varphi) }\,e^{-i\psi} \hat{\sigma}_-\right] + \text{c.c.} , \nonumber
\end{align}
where as before $\ell \in Z $ are associated with an OAM of $\pm \ell \hbar$ per photon. Such beams can be experimentally generated by transmitting radially polarized light through a quarter wave plate \cite{lerman2010}. We note that these light beams, just like radially polarized light featured in Fig.~\ref{fig:concurrence}d) have maximal concurrence, however their interaction with our atomic state interferometer differs. The interaction of atoms with hybrid polarized light can be interpreted as a coupling of sinusoidally varying light amplitudes with different magnetic sublevels $\ket{g_{\pm}}\rightarrow \ket{e} $. This can be useful for a polarization dependent measurement in an atomic system \cite{Fatemi2011,Luo2020}. Here we show the results for $|\ell|=1$ light beams, although the core results hold for higher $|\ell|$ values also.
The beam structures and corresponding absorption patterns for a selection of such hybrid vector vortex beams are shown in Fig.~\ref{fig:hyb concurrence}. Note that for Fig.~\ref{fig:hyb concurrence}a), light along the x and y axis of the beam is parallel and perpendicular to the magnetic field respectively, allowing the development of dark states, whereas for c) the linear polarization is at an angle of $\pm 45$ degrees to $\vec{B}$, with intermediate values taken in b).  

Not surprisingly, the Zeeman shifts of the partial dressed states vary across the beam profile in response to the alternation between right and left circular polarized beam areas,
\begin{align}
\label{eq:energy_hybrid_concurrence}
E_c &= -\hbar\Omega_L\cos\theta_B \cos{2\varphi},\nonumber \\
    E_g &= -\hbar\Omega_L\frac{\cos\theta_B\sin{4\varphi}}{2 M^2} \Big(\cos^2\theta_B  \sin{2\varphi}\\
&\hspace {4cm}+\sin^2\theta_B\cos2\psi^{\prime}\Big),\nonumber\\
    E_d &= -\hbar\Omega_L\frac{\cos\theta_B\sin^2\theta_B}{2 M^2} \cos{2\varphi} \big({1}-{\sin{2\varphi} }\cos2\psi^{\prime}\big). \nonumber
    \end{align}
The corresponding hopping rate between gray and dark state is
\begin{align}
\label{Eq:jgd_con}
        \mathcal{J}_{gd} 
    &=-\frac{\hbar\Omega_L}{M^2}\sin \theta_B \big[\bar{J} \sin {2\varphi} \cos^2 \theta_B \\
&\hspace{1.5cm}-\sin^2\theta_B (\cos{2\varphi} \cos{2\psi'}+i \sin{2\psi'})J/2\big], \nonumber 
\end{align}
with 
$$M^2=(1-\cos^2\theta_B\,\cos4\varphi+\cos2\psi^{\prime}\sin^2\theta_B \sin2\varphi)/2.$$  

In order to investigate the various different rotational symmetries exhibited by this system, we investigate the angular dependence of the transition rate Eq.~(\ref{Eq:P}) for a couple of cases, where the analytical form simplifies. For $\phi_B=0$ and $\psi=0$ (corresponding to Fig.~\ref{fig:hyb concurrence}a), we find
\begin{align}
P^\text{hyb}_{d\rightarrow e} \propto & \left[\frac{\big( -1+\sin2\varphi\big)}{-1+\cos^2\theta_B \cos4\varphi-\sin^2\theta_B \sin2\varphi}\right] \\
&\quad\times\big(4 \sin^3 \theta_B + (5 \sin\theta_B +\sin3\theta_B) \sin2\varphi \big)^2.\nonumber
\end{align} 
The symmetry of the absorption pattern depends on the magnetic field inclination $\theta_B$. 
At $\theta_B=\pi/2$, for instance, we find $P^\text{hyb}_{d\rightarrow e} \propto \cos^2 2\varphi$, indicating a 4-fold symmetry.
At $\theta_B=\pi/3$, however, the expression becomes
\begin{align}
P^\text{hyb}_{d\rightarrow e} \propto \frac{\,\big(-1+\sin2\varphi \big) \big( 3+5\sin2\varphi\big)^2}{-4+\cos 4\varphi-3\sin2\varphi},
\end{align} 
corresponding to a 6-fold symmetry in the absorption profile - features that could be investigated more generally by analyzing the angular Fourier series of the absorption patterns. 

We also note that, as the transverse component of the magnetic field increases, the visibility of the interference fringes also increases. For a radially hybrid polarized beam ($ 2\psi=0$), the visibility of interference fringes is maximum. As $2\psi$ increases, the beam becomes a `swirly' hybrid polarized beam, showing reduction in the fringe visibility. At $2\psi=\pi/2$, the absorption fringes disappear completely when the magnetic field is completely transverse. Under these circumstances, the system remains no longer spatially phase sensitive.

\begin{figure*}[th]
   \centering
\includegraphics[width=0.97\linewidth]{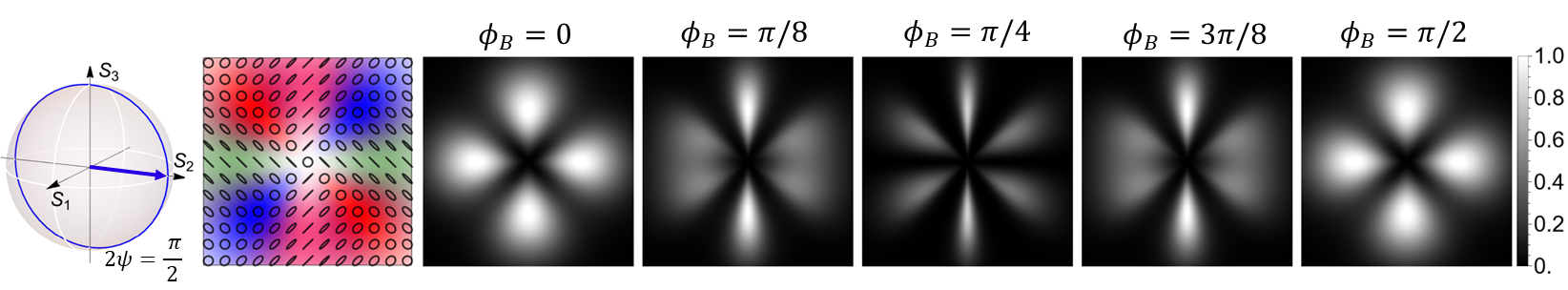}
    \caption{
    Absorption patterns for the hybrid vortex beam of Fig.~\ref{fig:hyb concurrence}
    for a fixed magnetic field inclination $\theta_B=\pi/3$, with $\phi_B$ rotating from $0\, \mathrm{to} \,\pi/2$. Due to the asymmetry of the polarization profile, a rotation of the magnetic field no longer corresponds to a rotation of the absorption pattern.}
    \label{fig:hybrid_varing_phib}
\end{figure*}

Finally, we turn our attention to a rotation of the magnetic field around the propagation axis, i.e.~a variation of its azimuthal angle $\phi_B$. We have noted early on in Eq.~(\ref{eq_phiB}), that the atomic dynamics are determined by the difference between the  
orientation of the polarization ellipse $\psi$ with respect to the magnetic field $\phi_B$. For the rotationally symmetric light profiles considered in Fig.~\ref{fig:concurrence}, a rotation of the magnetic field around $\phi_B$ results in a proportional rotation of the absorption pattern. For the hybrid polarization vortices considered here, however, this is no longer true, as shown in Fig.~\ref{fig:hybrid_varing_phib}, where a magnetic field rotation relates to a modification of the symmetry of the absorption pattern.

\subsection{Optical skyrmions}
\begin{figure*}[ht!]
\centering
\includegraphics[width=1.0\linewidth]{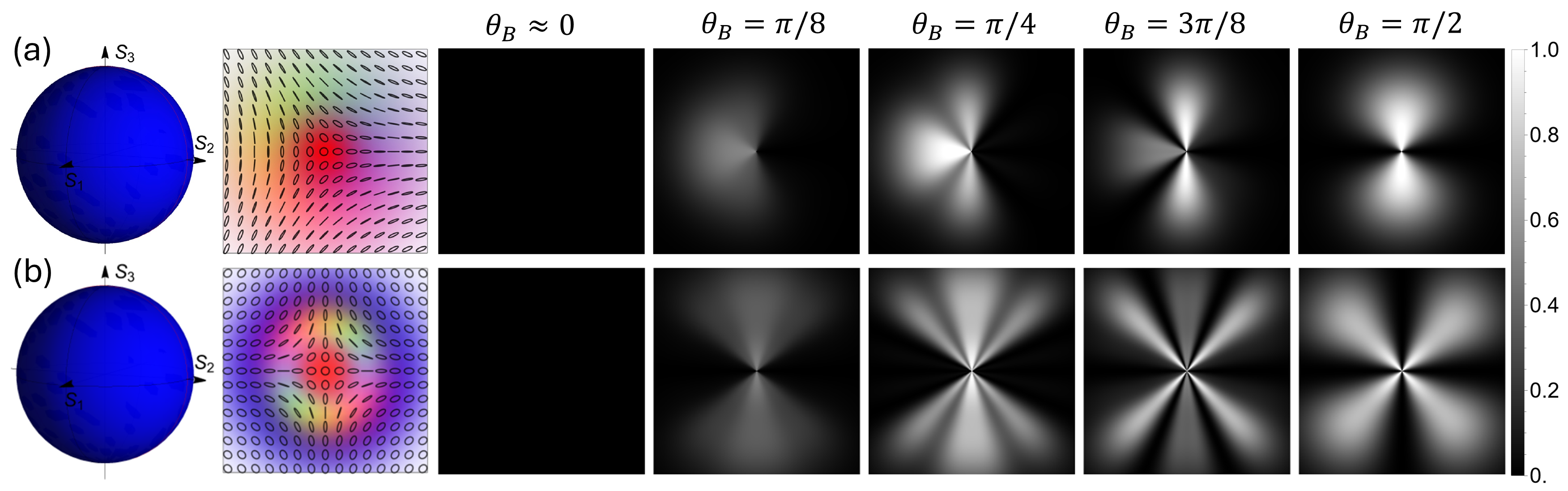}
 \caption{Asymmetric absorption patterns for skyrmionic beam\rev{s with skyrmion numbers 1 (a) and 2 (b)} generated \rev{according to Eq.~\ref{eq:poincare_beam} and Eq.~\ref{eq:poincare_beam2}, respectively. Note that figure a) shows } 
only the central part of the beam where intensities are sufficiently high. The magnetic field direction is changing \rev{in the} xz plane \rev{with $\theta_B$ increasing} from 0 to $\pi/2$. In the presence of a purely transverse magnetic field the transition rates for diagonally and anti-diagonally components are maximized whereas horizontal and vertical polarized light exhibit negligible interaction.}
    \label{fig:abs_poincare_beam}
\end{figure*}
In this section we investigate atom state interferometers driven by optical skyrmions, or Poincar\'e beams. These constitute a special class of vector light beams with spatially varying polarization distributions which cover the entire Poincar\'e sphere. 

Recently, optical skyrmions have been investigated in the context of their topology as well as various application in optical manipulation and optical communication \cite{Beckley:10,Xue2018,Galvez2021,McWilliam2023}. Very recently, optical skyrmions have also been investigated with respect to applications  in atomic magnetometers both theoretically \cite{5r63-5hl1} in first experiments \cite{10018304,Tian2024}. 
In the context of this paper, they allow us to test light matter interaction for all possible polarization states simultaneously, recovering the full dynamics as illustrated in Fig.~\ref{fig:Transitions_combined}. 

The simplest optical skyrmions (with skyrmion number 1) can be generated as a superposition of two $\rm{LG}$ modes having topological charges of 0 and 1 encoded onto their left and right circular polarization,
\begin{align}
\label{eq:poincare_beam}
\vec{E}_\text{S\rev{1}}(\vec{r}_{\bot})
&\propto\mathrm{LG}_{0}^{0} \,\hat{\sigma}_{-}+\mathrm{LG}_{0}^{1}\,\hat{\sigma}_{+} .
\end{align}
\rev{In a similar way, a polarization texture with skyrmion number 2 can be obtained as weighted superimposition 
\begin{align}
\label{eq:poincare_beam2}
\vec{E}_\text{S2}(\vec{r}_{\bot})
&\propto 2 \mathrm{LG}_{0}^{0} \,\hat{\sigma}_{-}+\mathrm{LG}_{0}^{2}\,\hat{\sigma}_{+},
\end{align} 
where the weighting factor was chosen to ensure a more balanced coverage of the Poincar\'e sphere.}
While the transition probabilities are readily calculable, their analytical form is sufficiently complicated to not be directly illuminating, and we will omit it here. Nonetheless, we note that the polarization structure in the beam profile is entirely asymmetric. It is therefore not surprising that the subsequent absorption patterns also show some asymmetric behaviors, as illustrated in Fig.~\ref{fig:abs_poincare_beam} \rev{a)}. We note that, while the polarization profile of the light given in Eq.~\eqref{eq:poincare_beam} contains all polarizations, its (right hand polarized) center is much brighter than the (left hand polarized) outer areas. In our figure we have `zoomed' in on the brighter inner region. \rev{The beam displayed in Fig.~\ref{fig:abs_poincare_beam} b) was chosen to display all polarization directions at reasonable light intensities, hence allowing us to test the interaction of generic vector light with the atomic state interferometer.}
As explained earlier, for a magnetic field along the propagation direction ($\theta_B=0$) the absorption vanishes. Tilting the beam in any direction results in increased absorption along the tilt direction, which changes in structure as the tilt increases. Once the magnetic field is purely transverse to the propagation direction, dark states develop where minor or major axis of the polarization ellipse is aligned with the magnetic field, resulting in two-fold absorption patterns at $\theta_B=\pi/2$. These geometric considerations explain, why optical skyrmions may be particularly beneficial for atom magnetometry.
We finally note, that our results agree qualitatively with predictions in \cite{5r63-5hl1} based on numerical evaluation via Liouville equations .

\subsection{HG beam}
We have, so far, considered only rotationally symmetric vector light, which can be encoded in LG modes and circular polarization states, making such light particularly suited to polar coordinate systems. Our formalism, however, holds for generic beams and in this final section we illustrate this by investigating optical vector beams created by superpositions of higher order Hermite-Gaussian $(\mathrm{HG})$ modes. The symmetry of these are best understood in linear polarizations, and a Cartesian coordinate system. We define the electric field as
\begin{align}
\label{eq:HG beam}
\vec{E}_n(\vec{r}_{\bot})\propto\mathrm{HG}_{0 n}\hat{x}+\mathrm{HG}_{n0}\hat{y}\,,
\end{align}
where $n\in{\mathbb{Z}^{+}}$.
This superposition of modes was chosen to once again form a radial beam at $n=1$ as a reference point, however higher order modes have a spatial intensity structure in addition to their polarization distribution. 

\begin{figure}[t]
\centering
\includegraphics[width=1.0\linewidth]
{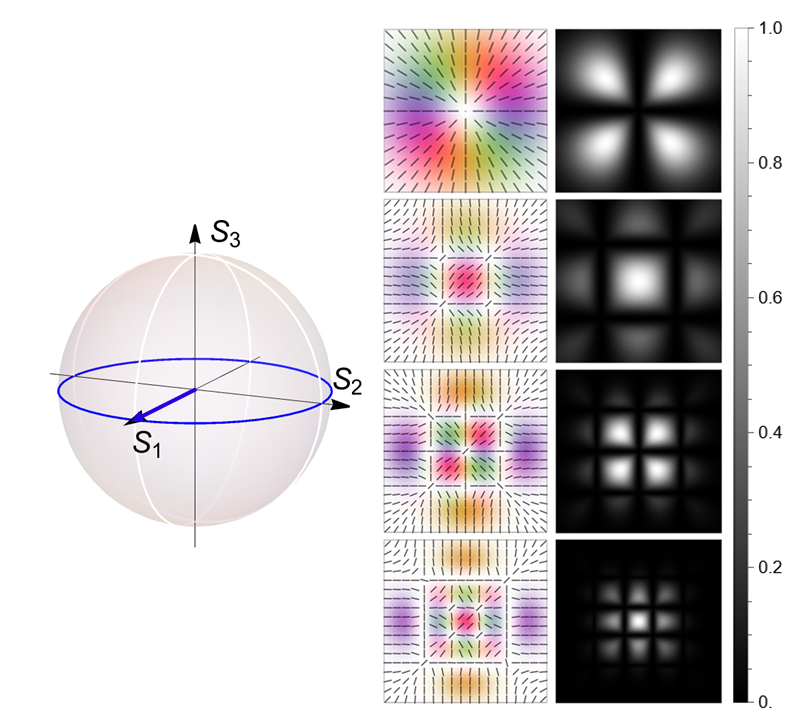}
 \caption{Absorption patterns for various polarization lattices, all containing solely polarizations along the equator of the Poincar\'e sphere. The middle column shows beam profile superposed with different HG modes (Eq.~\ref{eq:HG beam}) with $n=1$, 2, 3, 4 from top to bottom, The corresponding absorption patterns in a magnetic field $\vec{B}=B_0 \hat{x}$ are displayed in the right column.}
\label{fig:HG_beam_absorptions}
\end{figure}
\begin{figure*}[t]
\centering
\includegraphics[width=0.9\linewidth]
{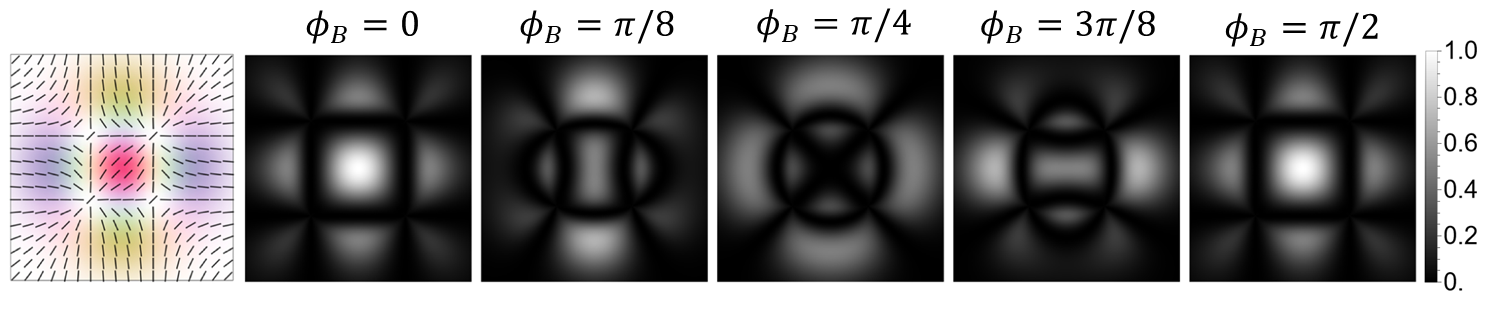}
 \caption{Absorption pattern for the higher order HG mode $\vec{E}_2$ in Eq.~(\eqref{eq:HG beam}), when the transverse magnetic field is rotated from $\phi_B=0$ to $\pi$/2.} 
\label{fig:HG_beam_abs_vaying_phib}
\end{figure*}

For simplicity, we consider a uniform magnetic field aligned solely in the plane transverse to the optical propagation i.e. $\theta_B=\pi/2$ and $\phi_B=0$. As with the Poincar\'e beam, we here omit the transition rate for the same reason. Fig.~\ref{fig:HG_beam_absorptions} shows beam profiles $(n=1,2,3$ and $4)$ and the corresponding absorption pattern. When $n=1$ it is a radial beam shows the similar absorption pattern as the extreme right of the Fig.~\ref{fig:concurrence}(d). Also for higher order modes, horizontal and vertical polarizations lead again to the formation of dark states, hence the light can only be absorbed in the regions where the two component beams overlap and we find an absorption profile with a square-lattice structure. It is possible that the resulting optical forces may be used as an additional parameter to tune novel optical lattices. 

Importantly, a rotation of the magnetic field does not result in a simple rotation of the absorption profile for these kinds of beams (as it would for the azimuthally symmetric beams displayed in Fig.~5).  Instead, a rotation of the magnetic field around the propagation axis manifests in a change to the absorption patterns. For the beam shown in Fig.~\ref{fig:HG_beam_abs_vaying_phib}, we see that the central absorption varies as a function of the magnetic field alignment. Furthermore, absorption appears prohibited along elliptical trajectories, whose eccentricity changes with magnetic field angle $\phi_B$. 

\section{Conclusions}

In this work, we have developed an analytical framework to describe atomic state interferometers driven by vector light with a generic spatial polarization structure. This extends the work of Refs.~\cite{Castellucci2021, Wang2024} and others and allows us to investigate the full parameter space of vectorial light-matter interaction for arbitrary complex vector beams without the need for lengthy numerical analysis. Specifically, we have shown that the interaction between the different excitation paths with an atomic state interferometer can be mapped to partially dressed states, including the spatial profile of dark states. 
We have demonstrated our approach for a wide range of vector beam structures, including radial polarized beams, hybrid vector beams, Skyrmion beams and higher order HG beams. The main benefit of our analytical method is that it allows us to obtain an intuitive understanding of the interplay between the polarization and the magnetic field, which can be hard to reach when relying on numerical simulations of optical Bloch equations or similar. Furthermore, these analytical results can be used to explore a wide range of quantum metrology applications, and identify vector light structures that optimize specific metrological tasks in, e.g. magnetometry applications. 

Naturally, our method comes with some limitations. Most importantly, we work in the regime where the natural decay of the excited state is much quicker than other dynamics, making long time dynamics inaccessible. We have not considered any Doppler broadening effects, making our results applicable for cold atomic samples, but with appropriate averaging methods also the behavior of warm atomic gasses could be described. 

Lastly, we have restricted our discussion to the calculation of spatially-dependent absorption coefficients, with the aim of understanding the interplay between atomic transparency, the magnetic field, and the structure of the vector light. Interestingly, the formation of spatially dependent dark states should also be associated with spatial dispersion patterns after propagation, which are worthy of investigation by themselves.
 We would like to investigate it along with the propagation of the light beam in an extended atomic cloud, to a future work.

\section*{Acknowledgements and Funding}
SJS, SF-A and NW acknowledge support through the QuantERA II Programme, with funding received via the EU H2020 research and innovation programme under Grant No. 101017733 and associated support from EPSRC under Grant No. EP/Z000513/1 (V-MAG). NW also acknowledges support from EPSRC via the grant no. EP/X033015/1, as well as support from the Royal Commission for the Exhibition of 1851.
KS acknowledges financial support from the UK Research and Innovation Council via grant EPSRC/DTP/W524359/1, and SJS via grant EPSRC/DTP/EP/W524359/1. 
We are grateful for discussions with Dr. Jinwen Wang for enlightening discussions on the interaction of hybrid vector beams as well as optical skyrmions with atom state interferometers.

\subsection*{Author contribution}

SF-A conceived the idea of this work and together with NW supervised the project. NW lead the theoretical discussions and wrote the appendix. KS wrote the code that generated the various absorption patterns, and produced all figures in Section III, SF-A generated those of Section II. All authors contributed to discussions, writing and editing of the manuscript. All authors have accepted responsibility for the entire content of this manuscript and approved its submission.

\subsection*{Conflict of interest}
The authors declare no conflict of interest.
\bibliography{bib}

\clearpage
\onecolumngrid
\appendix

\section{Transition probabilities and polarisabilities through perturbation theory}\label{app:pertTheory}
Given a Hamiltonian of the form Eq.~\eqref{eq:HamiltonianGeneric}, 
we are interested in calculating the transition probability, at some order in perturbation theory, of going from state $\ket{\psi_d}$ to $\ket{e}$. For notational simplicity, we will work in units of $\hbar = 1$ in this appendix. It is useful to note that we are in a rotating frame given by $\hat{U} = \exp\left(i\omega t \ket{e}\bra{e}\right)$, though this will not change anything for this particular calculation (but will if we calculate the atomic polarization). We are here interested in calculating $$P_{e\rightarrow d}~=~\left|\bra{e}\ket{\psi(t)}\right|^2.$$ To compute this, we first expand $\ket{\psi(t)}$ as
\begin{align}
    \ket{\psi(t)} = a_e(t)\ket{e}+a_c(t)\ket{\psi_c}+a_g(t)\ket{\psi_g}+a_d(t)\ket{\psi_d}.\nonumber
\end{align}
The Schr{\"o}dinger equation $i\partial_t \ket{\psi(t)} = \hat{H}\ket{\psi(t)}$ now simply yields the coupled equations
\begin{align}
    \frac{d a_e}{dt} &= -i\left[\delta a_e(t)+\mathcal{J}_{ec}a_c(t)\right], \nonumber\\
    \frac{d a_c}{dt} &= -i\left[E_c a_c(t) + \mathcal{J}_{cg}a_g(t) + \mathcal{J}_{ec}a_e(t)\right], \nonumber\\
    \frac{d a_g}{dt} &= -i\left[E_g a_g(t) + \mathcal{J}_{gd}a_d(t) + \mathcal{J}_{cg} a_c(t)\right], \nonumber\\
    \frac{d a_d}{dt} &= -i\left[E_d a_d(t) + \mathcal{J}_{gd}^*a_g(t)\right]. \nonumber
\end{align}
The formal solution to any one of these is
\begin{align}
    a_k(t) = a_k(0)e^{-i E_k t} - i e^{-i E_k t}\int_{-\infty}^t dt' e^{i E_k t'}J_{kj}a_j(t'),
\end{align}
where there can be multiple terms of the second kind. We are however interested in the situation where the dark states are populated, i.e. when $a_e(0) = 0 = a_c(0) = a_g(0)$ but $a_d(0) = 1$. To lowest order in perturbation theory, the contribution to $a_e(t)$ (which we need to calculate $P_{d\rightarrow e}$) is given by
\begin{align}
    a^{(3)}_e(t) = i e^{-i\delta t}\int_{0}^t dt' e^{i\delta t'}\mathcal{J}_{ec}\left[-i e^{-i E_c t'}\int_{0}^{t'} dt'' e^{i E_c t''}\mathcal{J}_{cg}\left(-i e^{-i E_g t''}\int_{0}^{t''} dt''' e^{i E_g t'''}\mathcal{J}_{gd}\left\{e^{-i E_d t'''}\right\}\right)\right],
\end{align}
where we should note that only the short time behaviour ($t\ll 1/\max(\delta,E_c,E_g,E_d)$) can faithfully be computed this way.
Thus, after some algebra and expanding to lowest order in the time $t$, we find
\begin{align}
    P_{d \rightarrow e} &= \left|a_e(t)\right|^2 \nonumber \\
    &= \frac{t^6}{36}\left|\mathcal{J}_{ec} \mathcal{J}_{cg} \mathcal{J}_{gd}\right|^2.
\end{align}
It is noteworthy that the transition rate, $P_{d \rightarrow e}/t \sim t^5$, grows quickly in time, ensuring that the transition happens quickly. A more realistic model would include the lifetime of the states. However, the quickest timescale is governed by the natural lifetime ($\Gamma$) of the excited state. The corresponding decay frequency is typically significantly larger than the Rabi and Larmor frequencies, $\Omega_R$ and $\Omega_L$, respectively, and therefore also the energies $E_c$, $E_e$, $E_d$ and the hopping rates $\mathcal{J}_{ec}$, $\mathcal{J}_{cg}$, $\mathcal{J}_{gd}$. Physically, this means that we only need to consider the short time dynamics of transitioning between the dark state and the excited state, as once the atom is in the excited state, it very quickly and incoherently decays and repopulates the ground states. In other words, we only need to capture times $t \leq 1/\Gamma$, which is done sufficiently well by perturbation theory. The maximum transition probability is therefore given by $t=1/\Gamma$, yielding
\begin{align}
    P_{d \rightarrow e} &\simeq \frac{1}{36\Gamma^6}\left|\mathcal{J}_{ec} \mathcal{J}_{cg} \mathcal{J}_{gd}\right|^2 = \frac{\Omega_R^2}{432\Gamma^6}\left|\mathcal{J}_{cg} \mathcal{J}_{gd}\right|^2,
\end{align}
as stated in the main text (after factors of $\hbar$ have been reintroduced).
\end{document}